\documentclass[twocolumn,notitlepage,nofootinbib,floatfix,superscriptaddress]{revtex4-2}
\usepackage{amsmath, amsfonts, amsthm, amssymb, bbm}
\usepackage[margin=1in]{geometry}
\usepackage{algorithm,algpseudocode}
\usepackage{graphicx,xcolor,tikz}
\usepackage[colorlinks=true, hyperindex, breaklinks, linkcolor=blue, urlcolor=blue, citecolor=blue]{hyperref}
\usepackage{physics}
\usepackage[utf8]{inputenc}
\usepackage{qcircuit}
\usepackage[utf8]{inputenc}
\usepackage{amsmath, amssymb}
\usepackage{geometry}
\usepackage{booktabs}
\usepackage{longtable}
\usepackage{listings}
\usepackage{algorithm}
\usepackage[percent]{overpic}
\usepackage{braket}
\usepackage{algpseudocode}
\newcommand{\pres}{p_{\text{res}}}

\algblockdefx[ForEach]{ForEach}{EndForEach}[1]{\textbf{for each} #1 \textbf{do}}{\textbf{end for each}}

\begin{document}

\preprint{APS/123-QED}

\title{Fault-tolerant distributed quantum computing with a single nucleus per node}

 \author{Yotam Vaknin}
 \email{yotam.vaknin.qc@gmail.com}
 \affiliation{Racah Institute of Physics, The Hebrew University of Jerusalem, Jerusalem 91904, Givat Ram, Israel}
 \author{Shoham Jacoby}
 \affiliation{Racah Institute of Physics, The Hebrew University of Jerusalem, Jerusalem 91904, Givat Ram, Israel}
 \author{Roi Nevo}
 \affiliation{Racah Institute of Physics, The Hebrew University of Jerusalem, Jerusalem 91904, Givat Ram, Israel}
 \author{Aleksander Kubica}
 \affiliation{Yale Quantum Institute \& Department of Applied Physics, Yale University, New Haven, CT, USA}
 \author{Alex Retzker}
 \affiliation{Racah Institute of Physics, The Hebrew University of Jerusalem, Jerusalem 91904, Givat Ram, Israel}
 \affiliation{%
AWS Center for Quantum Computing, Pasadena, California 91125, USA
}%

\date{\today}

\begin{abstract}

Distributed quantum computing interconnects small, high-quality nodes through optical links, but this architecture carries a pronounced asymmetry: in-node gates and measurements are cheap and high-fidelity, whereas inter-node communication relies on a low-coherence communication qubit and faulty photonics. Previous approaches overcame the noisy link by placing several high-quality data qubits in each node and consuming them for Bell pair and GHZ state distillation. Here we show that distillation can be avoided altogether. The key observation is that we can engineer a communication error bias, where photonic Bell pairs suffer frequent phase errors but only rare bit-flip errors. We design the syndrome-extraction circuits so that this phase noise appears solely as a measurement error that does not propagate to the data qubits, and is therefore suppressed by simply repeating the measurement; letting the error-correcting code itself, rather than a dedicated distillation subroutine, to purify the link.
This dramatically reduces the need for ancillary nuclei: Floquet codes require only a single data qubit per node, while general stabilizer codes require just one additional ancilla.
We demonstrate high error-correction thresholds throughout this regime, and we identify lattice surgery as inherently robust for this setting, enabling logical operations at a threshold close to that of quantum memory. As a result, the performance of the quantum computer is limited by the high-quality data qubits, while the requirements on photon indistinguishability and coherence of the communication qubit are substantially relaxed.

\end{abstract}
\maketitle

\section{Introduction}

The challenge of scaling quantum computers can be addressed through distributed
quantum computing. In these architectures, small quantum nodes,
each comprising a very limited number of qubits, are interconnected via optical
coupling for Bell pair generation. Numerous proposals along these lines have been put
forward~\cite{de_Bone_2024, Nickerson_2014, Sutcliffe_2025,
sunami2026entanglementboostinglowvolumelogical, Chandra_2025,Chandra_color_2025, ruskuc2025multiplexed,saggio2025building,inc2024distributed,sohr2024quantum,barral2025review, ramette2023faulttolerantconnectionerrorcorrectedqubits,shaw2026networked}, spanning a wide
range of device sizes.

Many recent proposals for distributed computing architectures are based on units composed of a communication qubit, an
electron spin used to emit photons, together with a data qubit and a small number of ancilla qubits
all encoded in nuclear spins \cite{aubele2026piqcscalabledistributedquantum, roggors2026singlemoleculespinphotoninterface,van2012decoherence,Beukers2025Control,sohr2026quantumcommunicationnetworksdefects,astner2022vanadiumsiliconcarbidetelecomready,Evans_2018,Nguyen_2019}. In these systems the number of available strongly coupled nuclear spins
is extremely limited, typically leaving only one or two data and ancilla qubits per node.
The architecture is also characterized by a pronounced mismatch in coherence times
between communication and data qubits, stemming from the much larger
gyromagnetic ratio of the electron compared with that of the nuclei. 

In most cases, the qubit with the lower coherence determines the overall performance of a quantum processor. Recent work has demonstrated that, by employing specialized architectures that leverage noise bias and implementing carefully engineered gates that maintain this bias, one can construct logical qubits whose performance is governed by the more coherent qubit, effectively mitigating the detrimental impact of the noisier qubit~\cite{putterman2025hardware,zuk2024robust,mehta2025bias,bi2026untanglingqldpccodesbiased}. 

In this work, we introduce the concept of bias to the distributed quantum computing setting, \textit{achieving performance that is ultimately constrained solely by the more coherent qubit}, i.e., the data qubit. By doing so, we gain two key advantages:
(1) we eliminate the need for ancillary qubits, 
reducing the number of nuclear spins per node to just one;
(2) the Bell pair fidelity threshold is significantly increased,
substantially relaxing the requirements on photon indistinguishability.

\begin{figure*}[t]
    \centering
    \includegraphics[width=\linewidth]{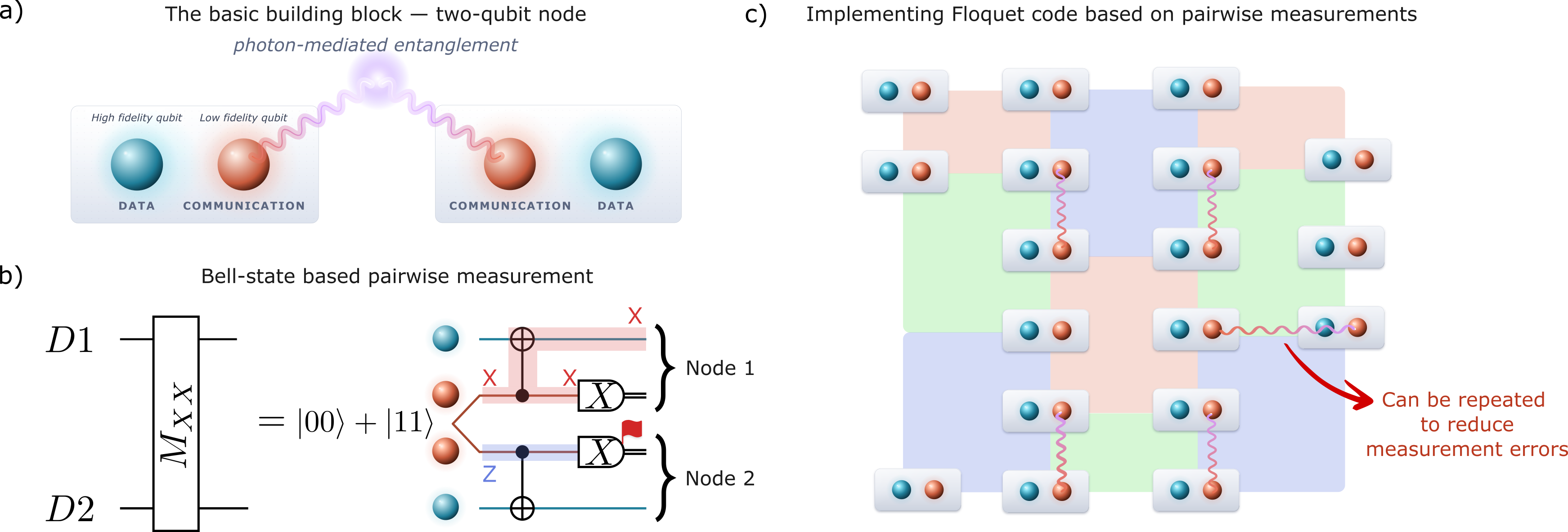}
    \caption{Using only one high fidelity qubit and one low fidelity qubit per node we can implement Floquet QEC protocols.
    a) The high fidelity qubit is used as the data qubit and the low fidelity qubit mediates interaction through optically generated entanglement.
    b) Using a Bell pair, one can implement non-local pairwise measurement on the data qubits. The large phase noise  on the Bell pairs flip the measurement result, while the lower rate bit-flip noise propagates to the data qubits. 
    c) By connecting multiple nodes through optical links, one can implement any dynamical code that is based on pairwise measurement. In this figure, we present the honeycomb code. The red, green and blue rectangles represent the three types of detectors of the code, each spanning six qubits.
    Other Floquet codes can be implemented in the same manner.
    }
    \label{fig:two-qubit-nodes}
\end{figure*}

Previous works have used post-selection to reduce the required fidelity. They used multiple
ancilla qubits and consume several noisy Bell pairs to distill a single
high-fidelity one~\cite{Nickerson_2014,debone2024thresholds}, followed by
combining multiple pairs to distill a high quality GHZ
state~\cite{Shor1996FaultTolerant,DiVincenzo2007SlowMeasurements,%
Tansuwannont2023AdaptiveSyndrome,Prabhu2023FaultTolerantSyndrome,%
DeBone2024DistributedSurfaceCode,Nickerson2013TopologicalNetwork,%
Nickerson_2014,debone2024thresholds,rodatz2026fault}. Naively, this would
suggest that distributed nodes require multiple nuclei to be useful at all:
distillation needs ancilla qubits to hold the intermediate states and to store
the accepted output, so a node reduced to a single data qubit has nothing to purify
with. The need for multiple nuclei is a major
drawback, as it immediately excludes a wide range of materials and defect types
and degrades gate fidelity, since every nucleus must be individually
controllable and spectrally distinguishable. 

Here, we skip that procedure and rely on QEC’s intrinsic noise-reduction mechanism, namely performing straightforward, repeated rounds of QEC syndrome measurements that do not propagate noise to the data qubits. This way, we avoid the ancilla qubits and post-selection required for distillation while effectively allowing the quantum code to \textit{distill} the correct syndrome measurement via repetition. 

The core idea is as follows. We use an inter-site Bell pair generation protocol that may carry substantial
phase noise (arising from distinguishable photons and short electron coherence) but only
rarely suffer bit-flip errors. We design the syndrome-extraction circuit so that
bit-flip errors propagate onto the data qubits, while phase errors only flip the syndrome measurement (or Pauli product) outcome. Because the
resulting measurement errors do not affect the data qubits, they can be
suppressed by simply repeating the measurement over several rounds.

We introduce an architecture for fault-tolerant quantum computing that makes full use of this bias, enabling a drastic reduction in the number of required ancilla qubits. Each computing node stores quantum information in a single
high-quality data qubit and uses the low-fidelity qubit for Bell pair generation
between different nodes (see Fig.~\ref{fig:two-qubit-nodes}). This minimal construction
already suffices to implement Floquet codes, which require only two-body Pauli
measurements, while retaining a high threshold and encoding rate. 

We also show that with only an additional data qubit per node, any stabilizer code can be implemented using measurements based on GHZ states~\cite{Shor1996FaultTolerant}. We show how the same bias can be utilized for this setting, recovering again the performance of the high-quality qubit. Our protocol still avoids the distillation of Bell pairs, and uses a single step of GHZ distillation, which is bias preserving.

We find that with our simple protocol, these codes can withstand a large amount of phase noise on the Bell pairs, thereby easing the stringent demands on photon indistinguishability. Our simulations show that the honeycomb code, which require only one nuclear spin,  has a threshold of up to $p_z \approx 11\%$, where $p_z$ is the phase-error probability of an individual inter-site Bell pair. 
For the surface-code memory, which requires two nuclear spins, the threshold for \emph{bulk-measurement errors}, which are the effective stabilizer errors that these
phase errors produce, exceeds $30\%$, corresponding to roughly $p_z \approx 10\%$ per Bell
pair. We further identify lattice surgery as an inherently
bulk-measurement-robust operation: for a residual error rate of $p_{\text{res}} = 10^{-3}$ on
the non-biased operations, its bulk-measurement threshold is approximately $15\%$.


\begin{figure*}[t]
    \centering
    \includegraphics[width=\linewidth]{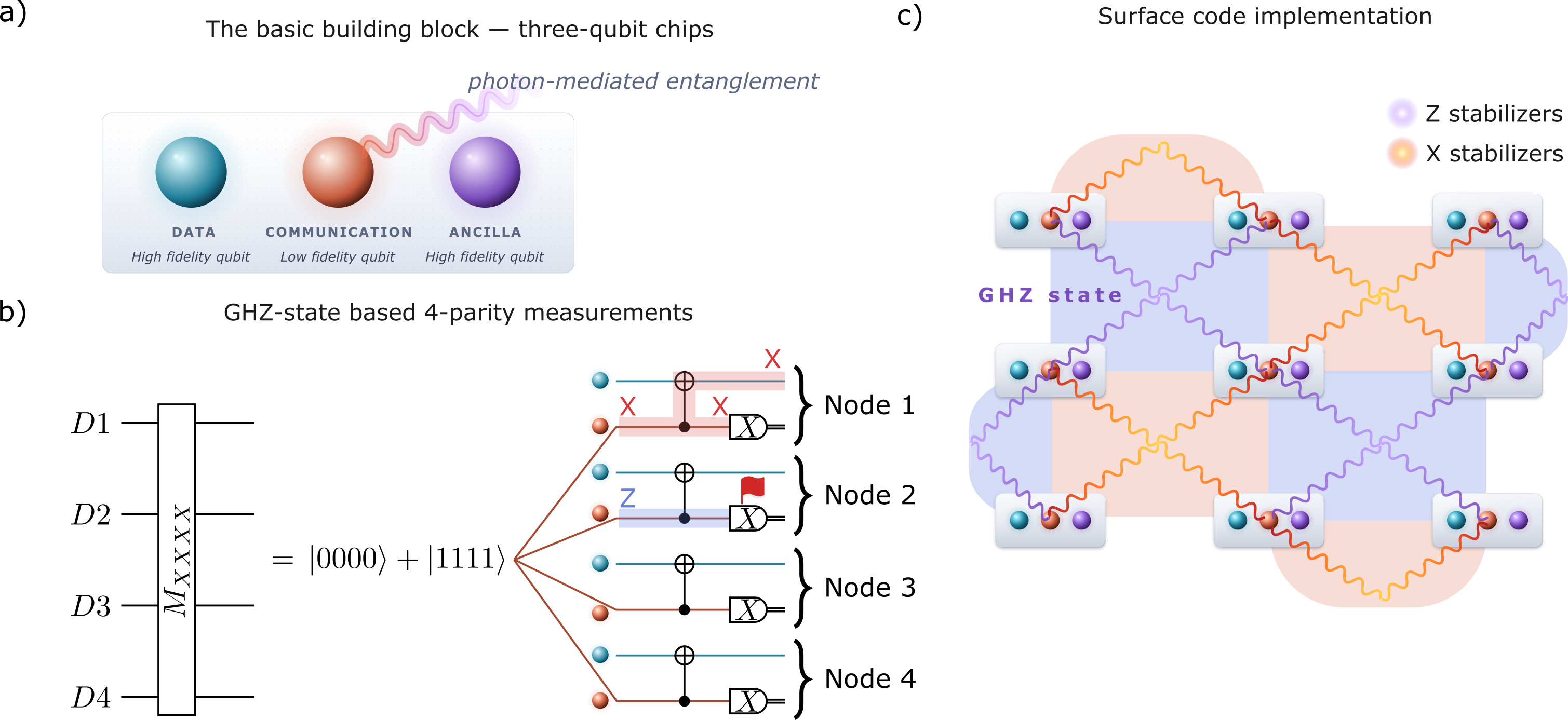}
    \caption{Adding a single high fidelity ancilla qubit per node enables the implementation of general stabilizer codes.
    a) Each node consists of a high fidelity data qubit, a low fidelity communication qubit that mediates interaction through optically generated entanglement, and a high fidelity ancilla qubit used for merging Bell pairs into GHZ states.
    b) Using a GHZ state distributed across four nodes, one can implement a non-local weight-four parity measurement on the data qubits. As in the two-body case, the dominant phase noise ($Z$) on the GHZ state only flips the measurement result, while the lower rate bit-flip noise ($X$) propagates to the data qubits.
    c) By connecting multiple nodes through optical links, GHZ states spanning the support of each stabilizer allow the implementation of any CSS code. In this figure, we present the surface code, but the scheme readily extends to other qLDPC codes~\cite{bravyi24b,panteleev22,kasai2026breaking}.
    \label{fig:three-qubit-nodes}}
    \label{fig:two-chip-floquet}
\end{figure*}

\section{Architecture}
\label{sec:architecture}

We consider distributed quantum nodes consisting of a low-coherence
communication qubit, typically an electron spin, together with one or more
higher-coherence qubits, usually nuclear spins. Since the coupling to photons can only occur through the electron spin, this spin is used to create Bell pairs with electrons at other nodes via linear optics, employing photon-based approaches such as the single-click \cite{cabrillo1999creation} and double-click \cite{barrett2005efficient} protocols (See appendix~\ref{app:dcp_error_model}).
The nuclei store the quantum information and serve as the data qubits or used as an ancillae for distillation. 

We study two settings, distinguished by the number of nuclei per node. With a
single nucleus, each node holds one data qubit, and we implement a Floquet code,
whose stabilizers are measured entirely through two-body Pauli measurements  (Fig.~\ref{fig:two-qubit-nodes}). 
Each such
measurement uses one Bell pair: the two communication qubits sharing the pair
are each coupled to their local data qubit by a {CX/CY/CZ}, and measuring the
communication qubits in the X basis then measures the joint
two-body parity (Fig.~\ref{fig:two-qubit-nodes}b). With a second nucleus available
as an ancilla, Bell pairs can be combined into GHZ states, which enable the
higher-weight stabilizer measurements of general stabilizer codes, which we exemplify with the implementation and analysis of the surface
code.

We distinguish between phase errors and all remaining errors in our noise model. We assume that the dominant contribution is phase noise affecting the generated Bell pairs, along with dephasing of the electron accumulated during the Bell pair preparation interval and during any additional two-qubit gates. For convenience, we group the electron’s total phase-error contribution into a single rate, $p_z$. All other noise mechanisms are taken to be considerably weaker, either because they are reduced via post-selection in the double-click and GHZ-generation protocols (see Sections~\ref{sec:stab_codes},\ref{sec:molecular_spins}), because of the long nuclear coherence times, or because idling noise is strongly suppressed by an efficient mechanism that protects the nuclei from electron-induced noise, as in Ref.~\cite{cohen2017protecting,brunelle2026silicon}. We denote the strength of this residual noise by $p_\text{res}$.

We now present protocols for the Floquet code and 
stabilizer codes that exploit this bias.

\subsection{ Floquet Codes}
\label{sec:bb_floquet}

Floquet codes are a generalization of stabilizer codes, defined by a schedule of not-necessarily-commuting measurements~\cite{Hastings_2021,gu2025,jacoby2026stairway}. These codes can greatly ease the requirement on the hardware by allowing for low-weight stabilizer measurements, and specifically allow for codes generated from solely two-body pairwise parity measurements. In our biased construction, these codes offer a major advantage for distributed computing, as they require only a single nucleus.

The planar honeycomb code \cite{Hastings_2021, Gidney_2021, Gidney_honeycomb_2022} is a planar, local Floquet code that achieves performance comparable to the standard surface code \cite{dennis2002topological,Wang_2011}. As with stabilizer codes, there are extensions to higher-rate constructions, including the hyperbolic Floquet code \cite{Higgott_2024hyperbolic, Fahimniya_2025_hyperbolic} and, more recently, the Stairway Code \cite{jacoby2026stairway}. 
Because distributed architectures are inherently non-local, these codes can be straightforwardly realized on the distributed systems considered in this work.

As illustrated in Fig.~\ref{fig:two-qubit-nodes}, we can use nodes with as few as two qubits to implement the pairwise-measurement required for Floquet codes. Specifically, in Fig.~\ref{fig:two-qubit-nodes}(c) we present one time step of the planar Honeycomb code, where the pairwise measurements are realized using Bell pairs.

The large phase noise $p_z$ translates into a high probability of measurement error on the two-body operators. Since these errors affect only the measurement outcome and not the data qubits, they can be suppressed by simply repeating each measurement, allowing the correct result to be recovered with high probability. This repetition slightly increases the accumulated residual noise, but yields a disproportionately large gain in the tolerable phase noise.

To use repeated measurement for error-correction, each pairwise Pauli-Product measurement is repeated $r$ times and combined into a single bit. Two natural strategies for that combination are: (i) a \emph{local} combiner, e.g.\ majority vote, that collapses the $r$ raw outcomes into one effective bit before handing the result to the decoder; and (ii) a \emph{global} decoder that consumes every individual outcome together with the rest of the syndrome graph \footnote{
This is the familiar inner/outer distinction from concatenated codes: strategy (i) hides the inner repetition behind an effective channel, while (ii) lets the outer decoder exploit the inner structure directly. We focus on (i) and leave (ii) to future work which may further improve our results.}.
In the case of a local combiner, it is also possible to compress the raw outcomes into multi-bit ``soft information''    \cite{pattison2021improved} that the outer decoder can then utilize.

\subsection{Floquet Simulations}

\begin{figure}
    \centering
    \includegraphics[width=0.8\linewidth]{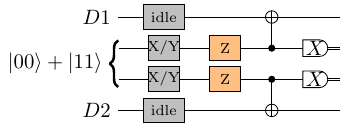}
    \caption{\textbf{Noise model for the Bell pair-based two-body Pauli
    measurement,} shown for a $X_1X_2$ measurement. A Bell pair of the two
    communication qubits is coupled to the data qubits by $\mathrm{CX}$ gates and the
    communication qubits are read out in the $X$ basis, yielding the joint $X_1X_2$ parity. Gray boxes are
    $p_\text{res}$-scale errors: the $X/Y$ dephasing of the Bell pair, data-qubit idling, and measurement errors. The orange $Z$ is approximately $p_{z}/2$ probability phase error on each communication qubit, originating from photon indistinguishability and lower coherence of the communication qubit. We set this value such that the phase-error per Bell pair is exactly $p_{z}$.
    The symmetric noise on the two qubits composing the Bell state is equivalent to a non-symmetric noise acting on only one of the qubits, which is the one we simulate.
    Given the high coherence of the single-nucleus data qubit, we don't follow the $\mathrm{CX}$ gate with a depolarizing channel. The dephasing error due to the low coherence of the communication qubit is absorbed as part of the error budget $p_z$, designated by the orange box.
    \label{fig:bell_noise_model}}
\end{figure}

We simulated the performance of the Honeycomb code using a noise model that can approximates our setting.  Figure~\ref{fig:bell_noise_model} shows the noise model for the Bell pair-based two-body Pauli measurement, illustrated for a $X_1X_2$ measurement. A Bell pair is generated on the two communication qubits; each is then coupled to its local data qubit by a CX gate, after which the communication qubits are read out in the $X$ basis to extract the joint $X_1X_2$ parity. 

The orange $Z$ box on each communication qubit labels the dominant $p_z$ phase error, which accounts for the entire phase-error-budget of the communication qubit (see appendix~\ref{app:QEC_noise_model} for  details on our noise model). The gray boxes denote $p_\text{res}$-scale noise: the $X/Y$ dephasing of the Bell pair, idling of the data qubits, and measurement errors. Given the high coherence of the single-nucleus nodes we treat the two-qubit-gate noise as negligible, which can be incorporated to leading order, in the error mechanisms already specified.

\begin{figure}
    \centering
    \includegraphics[width=\linewidth]{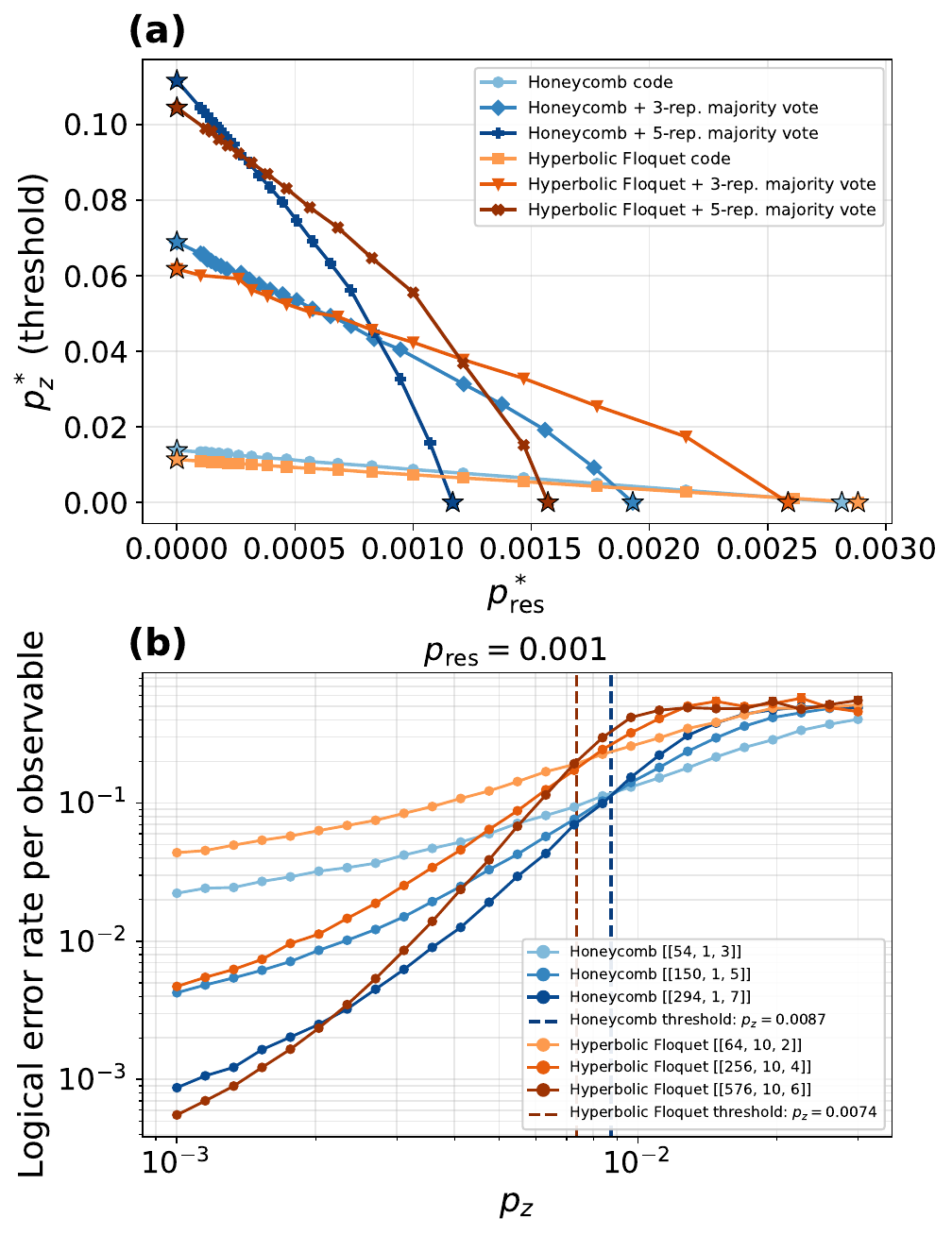}
    \caption{\textbf{Threshold performance of the Honeycomb and hyperbolic Floquet codes under the biased noise model, and the improvement from repeated measurements}.
    (a) Phase-noise threshold $p_z$ as a function of the residual error rate $\pres$. For both codes, combining three/five repeated pairwise measurements via majority vote substantially raises the $p_z$ threshold relative to a single measurement, at the cost of a modest reduction in the tolerable $\pres$. Stars represent the limiting cases of $\pres=0$ or $p_z=0$.
    (b)  Logical error rate per observable as a function of $p_z$ at fixed $\pres = 10^{-3}$, shown for increasing code distances of the Honeycomb code ($[[54,1,3]]$, $[[150,1,5]]$, $[[294,1,7]]$) and the hyperbolic Floquet code ($[[64,10,2]]$, $[[256,10,4]]$, $[[576,10,6]]$). The crossing of the curves marks the threshold, $p_z \approx 0.88\%$ for the Honeycomb code and $p_z \approx 0.74\%$ for the hyperbolic Floquet code (dashed lines).
    }
    \label{fig:3-shot-threshold}
\end{figure}

We simulated the memory performance of this code using stim \cite{gidney2021stim}, and used pymatching \cite{higgott2022pymatching} for decoding. The details of the simulation and decoding of the repeated measurement scheme are described in Appendix~\ref{SI:majority-vote-equivalent-noise-model}.  The result of our numerical simulation are shown in Fig~\ref{fig:3-shot-threshold}. We observe a threshold boundary as a function of $\pres$ and $p_z$, which define the correctable region. By repeating each measurement 3 or 5 times, we improve the $p_z$ threshold at the expense of $\pres$. 

The $\pres = 0$ points correspond to the measurement error threshold of the Floquet code $p_z^*$ for one repetition, and $p_i^*$ for $i$ repetitions.
Since for $\pres=0$ there is only measurement error induced by the Bell pair phase error, we can relate the 3-fold majority vote threshold to the $p_z^*$ threshold by $3 {p_3^{*}}^2(1-{p_3^{*}})  +{p_3^{*}}^3 = p_z^*.$ A similar condition applies to the 5-repetition case.

Naively, one would expect the $p_z=0$ points of single and 3- or 5-fold repeated measurements to have a $\pres$ threshold smaller by a factor of 3 or 5,
respectively, but we observe a different behavior in practice. Although the noise on the
data qubits is multiplied by 3 or 5, the syndrome data is much cleaner,
so a much larger fraction of the residual noise remains correctable. 
This is due to some error mechanisms controlled by $\pres$ becoming benign in the repeated-measurement scenarios.
We therefore find that for the hyperbolic Floquet code, when fixing $p_z=0$, the $\pres^*$ threshold is very close between a single measurement and 3-fold majority vote, around $\pres^*=0.25-0.28\%$.

Using the same noise model as above, we simulated the higher-rate semi-hyperbolic Floquet code \cite{Higgott_2024hyperbolic}. We used the $8.8.8$ tiling with $k=10$ and varied the inner-cell subdivision to obtain a family of codes at constant $k$. Fixing $\pres$ and scanning $p_z$, we find that the per observable logical-error-rate curves cross in a threshold-like manner, which defines the threshold boundary for the code. The results are shown in Fig.~\ref{fig:3-shot-threshold}. Applying the repeated-measurement noise model as well, we find an additional threshold boundary that behaves very similarly to the honeycomb code. Note that Floquet variants of higher-rate codes have already been constructed~\cite{jacoby2026stairway}, and it is expected that other high-rate codes could likewise be “Floquetified.”

\begin{figure}
    \centering
    \includegraphics[width=\linewidth]{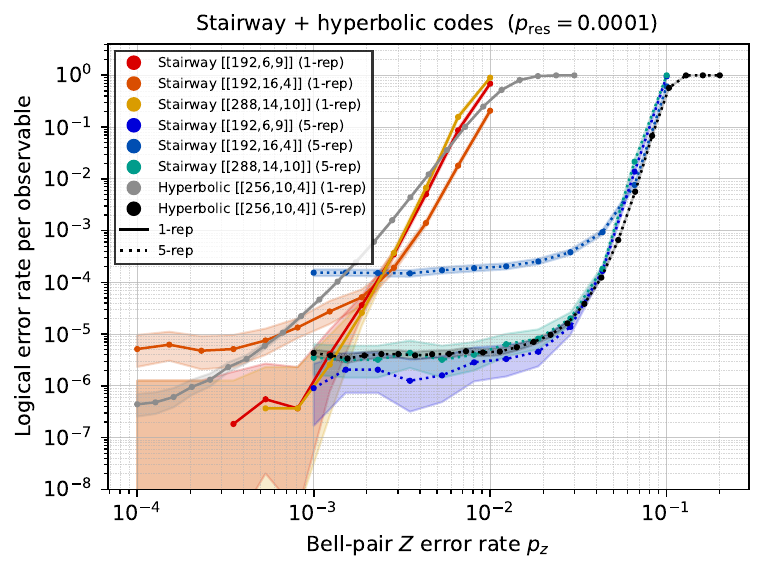}
    \caption{Logical error rates per $d$ rounds of the Stairway code family for 1-repetition (warm colors) and 5-repetition (cold colors) protocols. For comparison, we also plot the error rates of the [[256, 10, 4]] hyperbolic code under the 1-repetition (grey) and 5-repetition (black) protocols.
    }
    \label{fig:stairway-codes}
\end{figure}

We also simulated the Stairway code family~\cite{jacoby2026stairway} under the same noise model, demonstrating that our conclusion extends to these codes as well. In Fig.~\ref{fig:stairway-codes}, we present the logical error rate for a fixed $\pres=10^{-4}$. While Stairway codes exhibit a lower threshold, they achieve a lower error rate for similar block sizes in the sub-threshold regime. 
Since we can perform repeated measurements to increase the threshold for $p_z$ errors, these codes can be advantageous for systems with below-threshold $\pres$.






\subsection{Stabilizer Codes}
\label{sec:stab_codes}

\begin{figure}[b]
    \centering
    \begin{overpic}[width=\linewidth,]{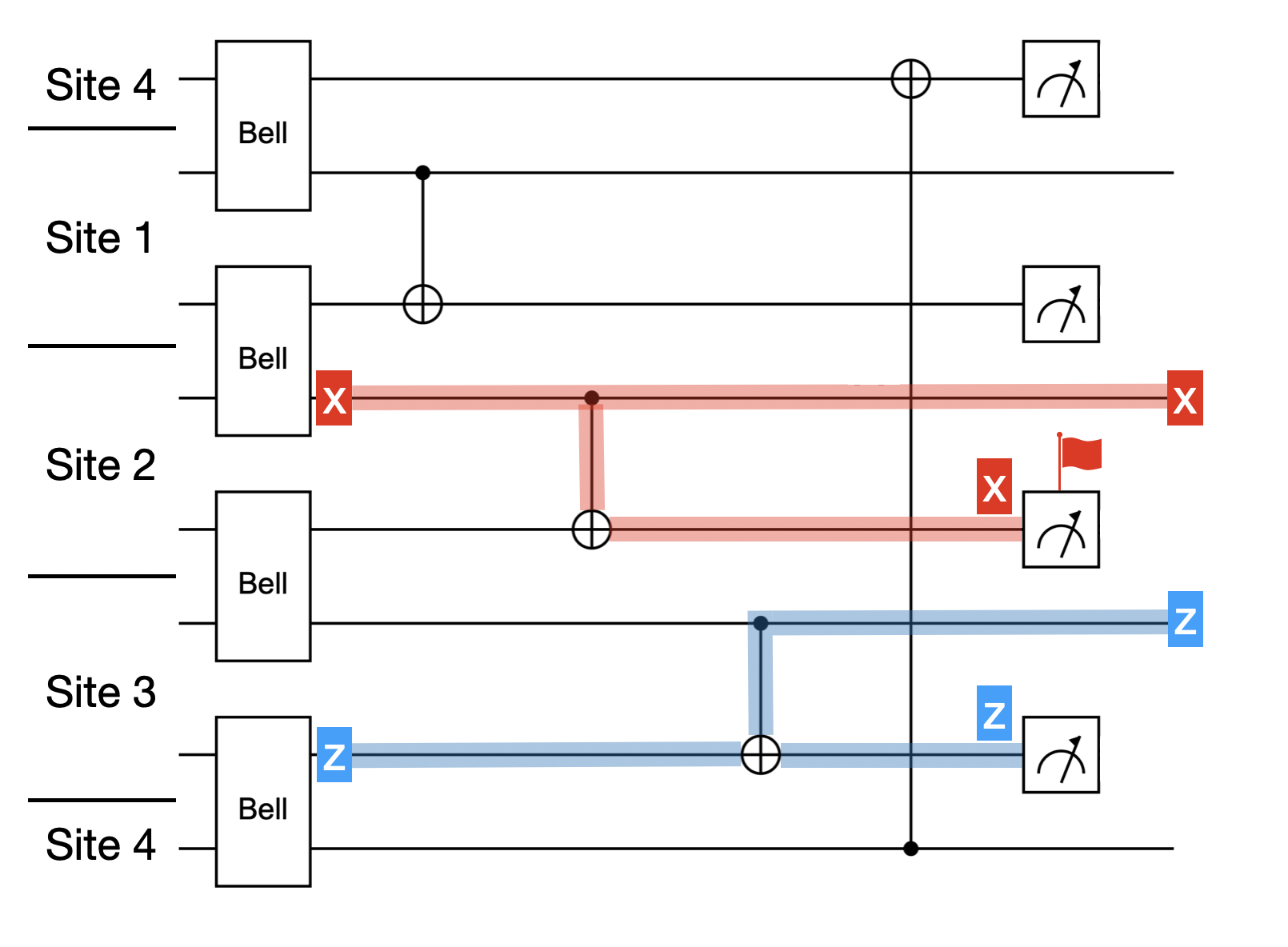}
    \end{overpic}
    \caption{
    A quantum circuit generating a GHZ state from four Bell pairs.
    The output state is a 4-body GHZ state split over 4 sites, where the two qubits in each site are the `communication' and `ancilla' qubits from Fig.~\ref{fig:three-qubit-nodes}. Site 4 is split for readability. All CX gates are internal to a common site. $X$ and $Z$ errors in the Bell pair generation process are followed through the circuit, propagating as $X$ and $Z$ errors on the GHZ state. Note that $Z$\,($X$) bias in the Bell pair generation process results in $Z$\,($X$) bias in the noise on the final GHZ state. A single $X$ error is always identified and can be post-selected over.}
    \label{fig:ghz_generation_protocol}
\end{figure}

With an additional ancilla qubit per node, two Bell pairs sharing a site can be
merged into a larger 3-qubit GHZ state. Additional Bell pairs can be similarly merged into a GHZ state of any size.
Such GHZ
states allow any stabilizer code to be implemented, since a stabilizer of any
weight can be measured through a sufficiently large GHZ state using Shor-style syndrome extraction~\cite{Shor1996FaultTolerant,delfosse2020shor_style}, see Fig.~\ref{fig:three-qubit-nodes}. 
While in the rest of the paper we concentrate on the surface code and thus GHZ states of four ancillas our results are quite general and could be readily applied for a variety of qLDPC stabilizer codes \cite{breuckmann21,gottesman14,tremblay22,panteleev22,leverrier22a,bravyi24b,cain2026shor,zhao2026towards,kasai2026breaking}, while utilizing larger GHZ states.
To
remain below threshold, however, we must preserve the same kind of bias as in
the two-body case, keeping the dominant phase noise from propagating onto the
data qubits. We now describe a protocol that achieves this and its impact on the QEC code.

The fault-tolerant generation of GHZ states for use in syndrome measurements has been thoroughly investigated \cite{Shor1996FaultTolerant,DiVincenzo2007SlowMeasurements,Tansuwannont2023AdaptiveSyndrome,Prabhu2023FaultTolerantSyndrome,DeBone2024DistributedSurfaceCode,Nickerson2013TopologicalNetwork,rodatz2026fault}, and in this work we adopt the following strategy.
The four-body GHZ state $\ket{\Psi} = \ket{0000} + \ket{1111}$ is generated
across four sites using four Bell pairs (Fig.~\ref{fig:ghz_generation_protocol}). The
protocol uses only \textsc{cnot} gates and $Z$-basis measurements, and is
bias-preserving in the sense that $X$/$Z$ errors on the input Bell pairs
propagate as $X$/$Z$ errors on the resulting GHZ state. It is not fully
deterministic: it produces $\ket{\Psi}$ up to single-qubit Pauli frame corrections,
which are inferred from the measurement outcomes (See Appendix~\ref{SI:GHZ}). Moreover, not every combination
of outcomes is admissible, since a single $X$ error on the Bell pairs is always
detectable, rendering $\ket{\Psi}$ even more robust to $X$ errors. 

\begin{figure}
    \centering
    \includegraphics[width=\linewidth]{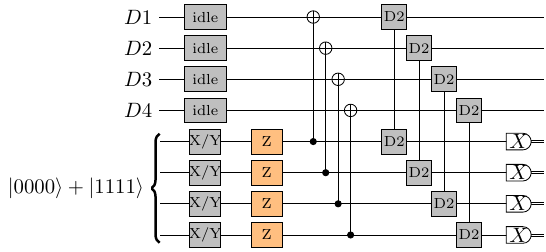}
    \caption{\textbf{Circuit-level noise model used in the surface-code
    simulations.} A weight-four $X$ stabilizer is measured by coupling four
    GHZ-entangled ancilla qubits (top) to four data qubits (bottom) via
    {CX} gates, followed by an $X$-basis readout (${H}$ and
    measurement). Gray boxes denote $p_\text{res}$-scale noise: data-qubit
    idling, single-qubit $X/Y$ errors, and the two-qubit depolarizing channels
    (D2) after each {CX}. The single orange $Z$ on each ancilla is the
    $p_z$ phase error; it captures both the electron's short coherence time and
    the biased Bell pair phase error, and represents the full error budget of the
    low-coherence qubit. Because it commutes through the {CX}, it only flips
    the measured syndrome, resulting in a bulk-measurement error.}
    \label{fig:ghz_noise_model}
\end{figure}

We model this effect by generating noisy Bell pairs with a $Z$ error of probability $p_z$ and an $X/Y$ residue error of probability $\pres$ on each qubit (See Fig.~\ref{fig:ghz_noise_model}).
We then simulate the protocol with noisy \textsc{cnot}s and noisy measurements,
where every \textsc{cnot} is followed by a two-qubit depolarizing channel with probability $\pres$, and every measurement outcome is classically flipped with probability $\pres$. As expected, we observe that the fidelity of $Z$-type stabilizers of the GHZ state are insensitive to $p_z$, whereas only the $X_1 X_2 X_3 X_4$ stabilizer degrades with increasing $p_z$; see Appendix~\ref{SI:GHZ} for further details.

Overall, our simulations demonstrate that the resulting noise  is well approximated by an independent $X$ error applied to each qubit composing the GHZ state with probability $\pres$ (since the additional $X$ noise on the Bell pairs is only second order). The post-selection rate of GHZ state is linear only in $\pres$, which we for $\pres=10^{-3}$ we observe to be below $1\%$; see Appendix~\ref{SI:GHZ} for more details.

Crucially, because the gates commute with the $Z$ error on the communication qubit, phase errors on the GHZ state never spread to other qubits and only flip the measured stabilizer value. We refer to this as a bulk-measurement error. It is distinct from a standard measurement error in that it occurs only in the bulk of the computation, never during the final readout, which is performed at higher fidelity by measuring the data qubits directly. Importantly, bulk measurement errors only produce time-like errors, as oppose to the final measurement that results in space-like errors. A memory experiment can only fail due to space-like error chains, making them highly robust to bulk-measurement errors. This can be seen by considering the limit $\pres \to 0$. In this limit $p_z$ has no effect on decoding: all intermediate syndrome data can be discarded, and the logical information recovered solely from the final (perfect) data-qubit measurement. Note that this benefit does not extend to Floquet codes, as in that case a memory experiment can fail with only bulk-measurement errors, which is why their $p_z$ threshold is lower.

\begin{figure}
    \centering
    \includegraphics[width=\linewidth]{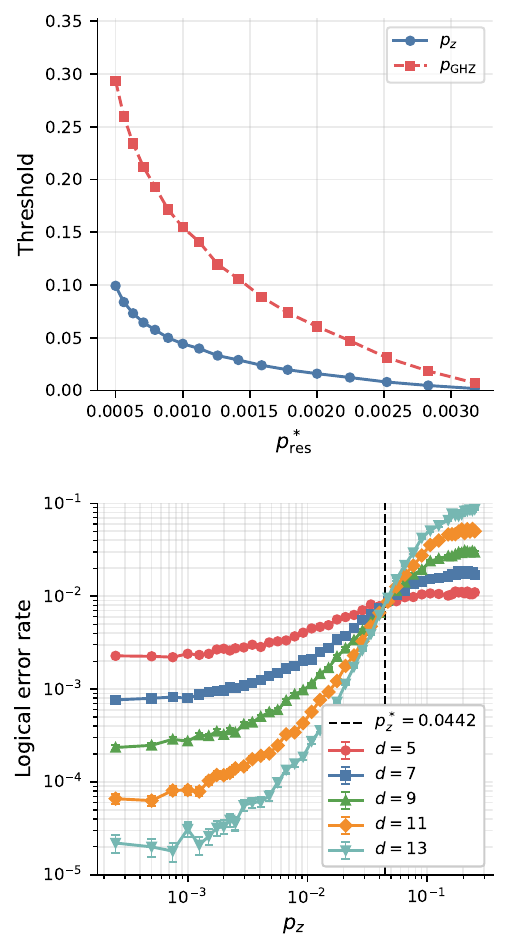}
    \caption{
  \textbf{Threshold behavior of the surface code memory under the
  bulk-measurement-error noise model.} (Top)~Threshold as a function of the
  residual operation error $\pres$, expressed in two equivalent ways. $p_z$ is
  the phase-error rate of an individual Bell pair, while $p_\text{GHZ}$ is the
  resulting effective error rate of a bulk stabilizer measurement: since each
  stabilizer is measured through a $4$-qubit GHZ state, the outcome is flipped
  whenever an odd number of its Bell pairs suffers a phase error, so
  $p_\text{GHZ} = 4p_z(1-p_z)^3 + 4p_z^3(1-p_z) \approx 4p_z$. Because syndrome
  extraction is performed via GHZ states, errors cannot propagate from a single
  ancilla to multiple data qubits, suppressing hook errors and other complex
  noise mechanisms and thereby substantially improving the $\pres$ threshold.
  (Bottom)~Representative threshold crossing at $\pres = 10^{-3}$, showing the
  logical error rate as a function of $p_z$ for code distances
  $d = 5, 7, 9, 11, 13$. The curves intersect at $p_z \approx 4.4\%$,
  corresponding to a bulk measurement error $p_\text{GHZ} \approx 15\%$
  }
    \label{fig:surface_code_threshold}
\end{figure}

\begin{figure*}
    \centering
    \includegraphics[width=\textwidth]{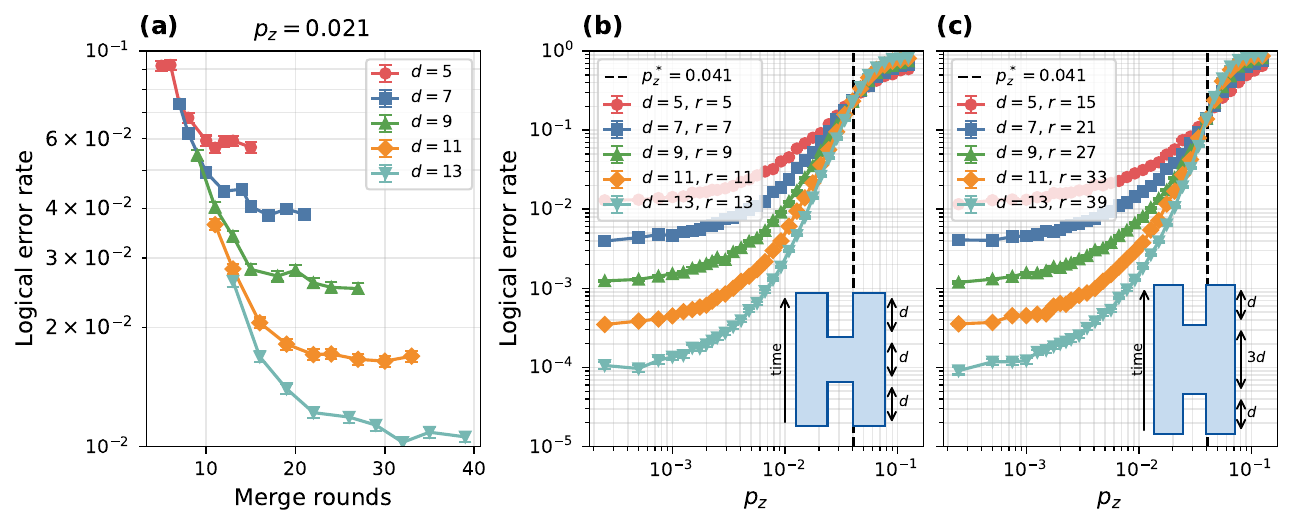}
    \caption{
  \textbf{Lattice surgery logical error rate}.
  We simulate the full two-patch lattice-surgery cycle: two distance-$d$ surface-code
  patches are each initialized in a logical $Z$ eigenstate and evolve independently for
  $d$ rounds, are merged for $r$ rounds to measure the joint parity $Z_1 Z_2$, and are
  then split and evolve independently for a further $d$ rounds. Three logical observables
  are tracked: the individual logical operators $Z_1$ and $Z_2$ of the two patches and
  the merge outcome $Z_1 Z_2$. A shot counts as a failure if any of them is
  decoded incorrectly, so the reported logical error rate is the error probability of the
  entire process.
  \textbf{(a)}~Logical error rate as a function of the number of merge rounds $r$ at
  $p_z = 0.021$ and fixed $p_{\rm res} = 10^{-3}$, for different code distances $d$.
  \textbf{(b,c)}~Logical error rate as a function of $p_z$ for merge rounds $r = d$ (b) and
  $r = 3d$ (c), at fixed $p_{\rm res} = 10^{-3}$. The dashed vertical line marks the
  threshold $p_z^* \approx 0.04$, which is essentially independent of the number of merge
  rounds. \emph{Insets:} spacetime schematic of the protocol (time running upward), showing
  the two code patches evolving separately for $d$ rounds, merging 
  for $r = d$ or $r = 3d$ rounds, and separating again for $d$ rounds.
  Away from threshold the two merge schedules (b) and (c) give nearly identical logical error
  rates, while close to threshold the difference becomes significant, as quantified in (a):
  the error rate drops sharply with the first few merge rounds and then saturates, so only a
  small number of additional rounds beyond $r = d$ is needed to reach the error-rate floor.
  }
    \label{fig:ls_paper}
\end{figure*}

An additional advantage of this approach concerns the effect of $X/Y$ errors on the GHZ state. Since the GHZ generation protocol we employ is fault tolerant \cite{Shor1996FaultTolerant,DiVincenzo2007SlowMeasurements,Tansuwannont2023AdaptiveSyndrome,Prabhu2023FaultTolerantSyndrome,DeBone2024DistributedSurfaceCode,Nickerson2013TopologicalNetwork}, weight-2 combinations of $X$ and $Y$  errors occur on the 4-body GHZ state only with probability $\mathcal{O}(\pres^2)$(See Appendix~\ref{SI:hook_errors}). Since only $X$ errors propagate outward to the data qubits, the weight-4 stabilizer measurements are fault tolerant and no single error spreads to multiple data qubits. 
We analyze the general case of weight-$w$ stabilizers in Appendix~\ref{SI:hook_errors}, and note here that the scheme remains favorable in that setting as well.

\subsection{Surface Code Memory}

As a concrete benchmark for the stabilizer code scheme, we simulated multiple rounds of QEC with the surface code using the generated GHZ states. The stabilizers are split into four rounds, such that every
data qubit lies in the support of at most one stabilizer per round. In each
round we generate a GHZ state across the nodes holding the data qubits in the
support of a given stabilizer, and use it to measure that stabilizer via the
circuit of Fig.~\ref{fig:two-chip-floquet}.
We repeat these stabilizer measurements for $d$ rounds, where $d$ is the distance of the code, and finally measure the data qubits directly to recover the logical information. 

The noise model we use is design to model the results of out GHZ simulation from Section~\ref{sec:stab_codes}.  We keep the convention that $p_z$ describes the phase-noise error budget of the communication qubit, and $\pres$ describes the fidelity of all other operations, including idling and 2-qubit gates.
Here, we  include 2-qubit gate errors, as they tend to be of lower fidelity than the single nucleus case. The fault-tolerance of the GHZ state means we only need to directly model weight-1 errors on the GHZ qubits. The noise model is showen in Fig~\ref{fig:ghz_noise_model}, and described in full details in Appendix~\ref{app:QEC_noise_model}.

The results of these simulations are shown in Fig.~\ref{fig:surface_code_threshold}. We see that at $\pres=10^{-3}$, the $p_z$ threshold is roughly $4.5\%$, while we can still tolerate $p_z>10\%$ when  $\pres<0.05\%$. As we explained earlier, $p_z$ can reach as high as $50\%$ and still be correctable, as long as the residual error level is low enough. 

\subsection{Lattice Surgery}

Because memory experiments are inherently robust to bulk measurement errors, a complementary experiment is necessary to demonstrate that our setting is useful for arbitrary quantum computation. Here we address the full case of multi-qubit computation by simulating a two-qubit gate using lattice surgery.  In Appendix~\ref{SI:stability} we also consider the more simplified case of a stability experiment~\cite{Gidney_stability_2022}, which directly probes the effect of long timelike error chains.
We show that lattice surgery is characterized by a threshold very similar to that of the memory experiment, which shows that the weak sensitivity to $p_z$ errors is not a particular property of memory experiments.  

Lattice
surgery~\cite{Horsman2012LatticeSurgery,Litinski2019GameOfSurfaceCodes,Vuillot2019GaugeFixing,Landahl2014ColorCodeSurgery} is a natural approach to multi-qubit logical operations in our setting. A lattice surgery measures a product of logical observables  by
temporarily introducing additional stabilizers and modifying the existing
stabilizers whose support overlaps the logical operators that are being measured.
The outcomes of these measurements reveal the value of the joint multi-qubit Pauli
product. Noisy syndromes corrupt this estimate, but repeating the measurement
over multiple rounds overcomes the noise, up to the point where the accumulated noise from additional rounds begin to degrade the logical qubits. 

We simulated lattice surgery between two surface code patches, to estimate the bulk-measurement error's effect on the information obtained by lattice surgery (Figure~\ref{fig:ls_paper}). We initialized the codes in the $\ket{0_L}$ state, and measured their joint Pauli product $Z_1 Z_2$ using lattice surgery, followed by individual $Z$ measurements on each patch ($Z_1$ and $Z_2$). The logical error rate is defined as the probability that any of these three Pauli outcomes is incorrect. Because of the symmetry in our idling errors, the dual
experiment in the $X$ basis would yield the same logical error rate.

The simulation sequence proceeds as follows: two patches are initialized in the $Z$ basis (all data qubits prepared in $\ket{0}$) and undergo $d$ rounds of GHZ based syndrome extraction. The patches are then merged into a single patch for a variable number of rounds $r\in [d,3d]$, after which they are split apart and each undergoes $d$ additional rounds of syndrome extraction before being measured in the $Z$ basis. The results are shown in Figure~\ref{fig:ls_paper}.

Remarkably, even in the lattice-surgery setting, the $p_z$ threshold remains essentially that of the memory case, approximately $4.5\%$ at $\pres=10^{-3}$. Although lattice surgery introduces many additional measurement errors, the logical fidelity plateaus after only a few additional merge rounds, beyond which further rounds bring no improvement. This behavior can be understood by separating the logical error into two contributions: a stability-type component, suppressed exponentially in the number of rounds, and a memory-type component, which grows only linearly. Far below threshold, the exponential suppression reaches the memory floor almost immediately, so that a handful of merge rounds already recovers the full memory performance. We make this decomposition precise using a stability-experiment simulation~\cite{Gidney_stability_2022}, described in Appendix~\ref{SI:stability}.

Transversal operations are naturally suited to distributed architectures, as they require only pairwise connectivity that can be established arbitrarily. 
In principle, one could generate Bell pairs between corresponding data qubits of two  codes and use them for gate teleportation, thereby realizing a transversal \textsc{cnot}. In practice, however, this approach is highly problematic: the Pauli corrections inherent to gate teleportation translate measurement errors into single-qubit errors on the data, producing a round in which the effective error rate scales as $p_z$ rather than $\pres$. This can be avoided by replacing the teleported gate with two-qubit
measurements that can be repeated. A \textsc{cnot} can be implemented by
combining an $XX$ and a $ZZ$ measurement, followed by single-qubit Pauli
corrections. This is the same technique used for lattice surgery between
logical qubits~\cite{fowler2019lowoverheadquantumcomputation}, applied here to
pairs of physical qubits. The joint measurements are themselves implemented
using Bell pairs, and, crucially, they can be repeated until the outcome is
sufficiently reliable before any correction is committed. A measurement error
therefore no longer propagates onto the data, and the effective error rate of
the round returns to $\pres$.
 We leave this analysis for future work.

\section{Biased Bell Pair without distillation}
\label{sec:molecular_spins}

Our protocol assumes that Bell pair generation yields a bias between phase noise and bit-flip noise. This bias can be achieved as a direct consequence of the Bell pair generation protocol, first described in Ref.~\cite{barrett2005efficient}, which we review here in the relevant detail for completeness.   We describe the protocol in more details, including leading noise terms, in Appendix~\ref{app:dcp_error_model}.

We will follow the double-click protocol \cite{barrett2005efficient, debone2024thresholds}. We refer to a single-\textit{click} as in instance where pairs of electrons are each excited to a superposition of the form $\ket{\phi}=\sqrt{1-p_1}\ket{0}+\sqrt{p_1}\ket{1}$, for some probability $p_1$. Then, both $\ket{1}$ states are excited to a photo-emitting level, and the photons are align to interfer through a beam splitter, with a photon-detector on each end. If a single photon is observed on either end, the electron pairs are projected onto a Bell pair $\ket{\Psi^\pm}=\ket{10}\pm\ket{01}$, the sign depending on the detected port. If the photon detector's number-resolving efficiency is low and there is substantial loss on the way, there remains some probability that the pairs are in the state $\ket{11}$, resulting in a bit-flip error. 

To further increase the bias, both electrons are flipped using a $\pi$-pulse, which stabilizes the Bell state but flips $\ket{11}$ to $\ket{00}$. Then, the two electrons are excited again to emit another photon. Given the initial state $\ket{11}$ the probability of observing a single photon again is, to first order, given by the probability of a dark-count, which is much smaller compared with the phase noise between the electron pair. This provides the noise bias for the \textit{double-click} Bell pair generation. This bias can be further increased by repeating the single-click case an even number of times.
This would exponentially supressing the bias up to the floor of the single-qubit gate errors.

\section{Discussion}

Distributed quantum computing is gaining traction as a promising architecture for fault-tolerant quantum computation, largely because of its potential to address the problem of scalability.  
Recent progress in the field has demonstrated a variety of defects and molecules that could serve as distributed nodes, making the efficient implementation of  QEC with a minimal number of ancilla qubits especially desirable. This is particularly important because the number of available data qubits in these systems is severely limited.

We showed that the requirements on the Bell pairs and low-fidelity
communication qubits can be made very lax: the tolerable phase noise can exceed $10\%$ with minimal or no distillation, for both Floquet codes and the surface code. For the surface code, we further showed that these results generalize to both quantum memory and logical computation.

Together, these results bring many existing distributed platforms within
reach using a simple protocol and a small number of nuclei. Importantly, we
showed that the seemingly trivial action of repeating a  measurement
can substantially improve the threshold, given the appropriate Floquet code.

Moreover, our scheme is also promising for higher-rate stabilizer codes,
where the stabilizer weight exceeds~$4$. In that regime avoiding hook errors is more challenging,  but we show that GHZ-based measurement can make them significantly rarer.
This may bring the circuit distance of these protocols to the code-capacity distance of the underlying qLDPC codes. We leave this direction for future work.

\begin{acknowledgements}
We thank Fernando Pastawski and Alon Salhov for helpful discussions.
A. R. acknowledges the support of the Israeli Science Foundation and the Directorate for Defense Research and Development (DDR\&D) (grant No. 675/24) and the Schwartzmann university chair.
\end{acknowledgements}

\section{Data Availability}

The scripts used in this paper, the stim circuits, and the sampled data are available in~\cite{vaknin2026data}.

\clearpage

\appendix

\section{Equivalent noise model of the majority-vote gadget}
\label{SI:majority-vote-equivalent-noise-model}

To suppress measurement errors in the pairwise measurement circuit we repeat the same pairwise stabilizer measurement $r$ times and combine the outcomes into a single bit.
Two natural strategies for that combination are: (i) a \emph{local} combiner, e.g.\ majority vote, that collapses the $r$ raw outcomes into one effective bit before handing the result to the decoder; and (ii) a \emph{global} decoder that consumes every individual outcome together with the rest of the syndrome graph.
This is the familiar inner/outer distinction from concatenated codes: strategy (i) hides the inner repetition behind an effective channel, while (ii) lets the outer decoder exploit the inner structure directly.
We focus on (i) and leave (ii) to future work.

We can model the majority-vote gadget as a single noisy pairwise measurement.
This reduction is not trivial: errors that occur in one repetition can propagate, through the shared data qubits and CX gates, into later repetitions, and the syndrome bits are then nonlinearly combined by the majority vote, so a naive product of single-shot channels misses these correlations.

\begin{figure*}[t]
    \centering
    \includegraphics[width=\linewidth]{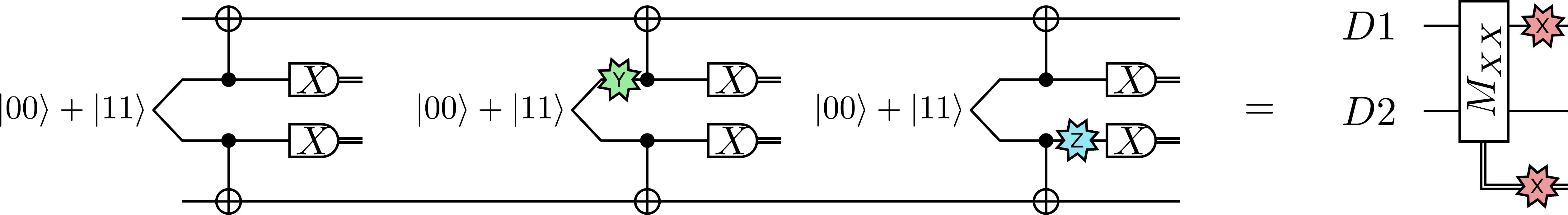}
    \caption{Performing pairwise measurement three times and taking majority vote can reduce measurement errors. Here, we show how two independent errors on a repeated circuit, when decoded using a inner majority decoder, can be equivalent to a single pairwise measurement with a correlated measurement error and a single-qubit Pauli $X$ error.
    }
    \label{fig:equivalent_majority}
\end{figure*}

Our approach is to start from the detector error model (DEM)\cite{gidney2021stim} of the repeated circuit, enumerate its error mechanisms up to leading order, fold the three syndrome detectors through majority vote, and read off an effective channel acting on (data, ancilla).
In Fig.~\ref{fig:equivalent_majority} we show an example of one error mechanism that is included in the enumeration, and the equivalent error mechanism it is replaced with (with corresponding probability).

The result is a list of weighted Pauli operations that, plugged into the outer simulator in place of a single pairwise measurement, reproduces the gadget's behavior to leading order in the physical error rate. In our numerical simulation, we only simulated the outer code directly, and applied to it the noise model induced by the repeated pair-wise measurements. 

\section{GHZ Generation Protocol}
\label{SI:GHZ}
\begin{figure*}
    \centering
    \includegraphics[width=\textwidth]{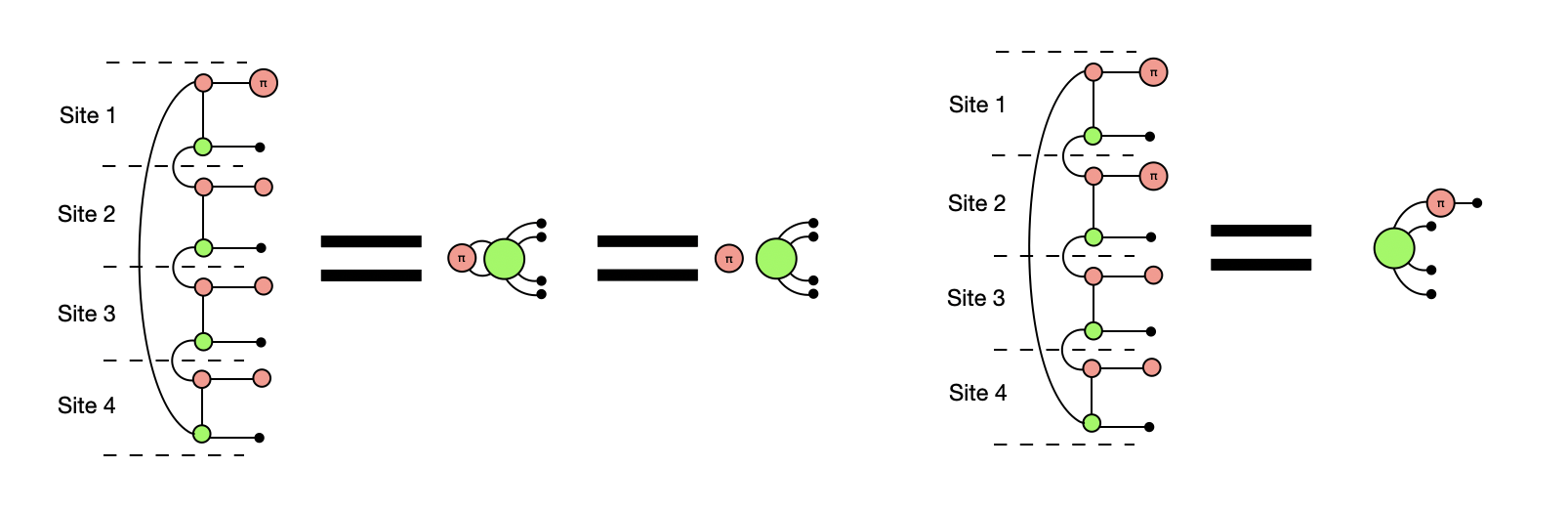}
    \caption{ZX-calculus picture of error detection and correction in the GHZ
    generation protocol. (Left) A single bit-flip error on an input Bell pair
    propagates to a $\pi$ on a measured leg, giving an odd-parity (rejected)
    outcome; the diagram reduces to zero and the error is detected. (Right) A
    pair of bit-flip errors reduces to a single residual $\pi$ on an output leg,
    i.e.\ a single-qubit Pauli correction on the GHZ state.}
    \label{fig:ghz_correction_and_detection}
\end{figure*}

The GHZ-generation protocol we employ~\cite{Nickerson_2014} fuses four Bell pairs into a
single four-body GHZ state. The corresponding circuit is shown in
Fig.~\ref{fig:ghz_generation_protocol}(a): a single layer of \textsc{cnot} gates is
applied, after which half of the qubits are measured in the $Z$ basis. The
individual measurement outcomes are random and specify the single-qubit Pauli
correction required to bring the output onto the ideal GHZ state (This
correction is purely classical and can be tracked in software). The explicit set
of corrections can be inspected in the accompanying \href{https://algassert.com/quirk#circuit=%7B%22cols%22%3A%5B%5B%22~95p7%22%2C1%2C%22~95p7%22%2C1%2C%22~95p7%22%2C1%2C%22~95p7%22%5D%2C%5B1%2C%22%E2%80%A2%22%2C%22X%22%5D%2C%5B1%2C1%2C1%2C%22%E2%80%A2%22%2C%22X%22%5D%2C%5B1%2C1%2C1%2C1%2C1%2C%22%E2%80%A2%22%2C%22X%22%5D%2C%5B%22X%22%2C1%2C1%2C1%2C1%2C1%2C1%2C%22%E2%80%A2%22%5D%2C%5B%22Measure%22%2C1%2C%22Measure%22%2C1%2C%22Measure%22%2C1%2C%22Measure%22%5D%2C%5B%22split7%22%5D%2C%5B%22%E2%80%A2%22%2C%22%E2%80%A2%22%2C%22%E2%97%A6%22%2C%22%E2%97%A6%22%2C%22X%22%5D%2C%5B%22%E2%80%A2%22%2C%22%E2%97%A6%22%2C%22%E2%80%A2%22%2C%22%E2%97%A6%22%2C%22X%22%2C%22X%22%5D%2C%5B%22%E2%80%A2%22%2C%22%E2%97%A6%22%2C%22%E2%97%A6%22%2C%22%E2%80%A2%22%2C%22X%22%2C%22X%22%2C%22X%22%5D%2C%5B%22%E2%80%A2%22%2C%22%E2%80%A2%22%2C%22%E2%80%A2%22%2C%22%E2%80%A2%22%2C%22X%22%2C1%2C%22X%22%5D%2C%5B%22%E2%97%A6%22%2C%22%E2%80%A2%22%2C%22%E2%80%A2%22%2C%22%E2%97%A6%22%2C1%2C%22X%22%5D%2C%5B%22%E2%97%A6%22%2C%22%E2%80%A2%22%2C%22%E2%97%A6%22%2C%22%E2%80%A2%22%2C1%2C%22X%22%2C%22X%22%5D%2C%5B%22%E2%97%A6%22%2C%22%E2%97%A6%22%2C%22%E2%80%A2%22%2C%22%E2%80%A2%22%2C1%2C1%2C%22X%22%5D%2C%5B1%2C1%2C1%2C1%2C%22Amps4%22%5D%5D%2C%22gates%22%3A%5B%7B%22id%22%3A%22~th5n%22%2C%22name%22%3A%22BELL%22%2C%22circuit%22%3A%7B%22cols%22%3A%5B%5B%22H%22%5D%2C%5B%22%E2%80%A2%22%2C%22X%22%5D%2C%5B1%2C%22%E2%80%A2%22%2C%22X%22%5D%2C%5B1%2C1%2C1%2C%22%E2%80%A2%22%2C%22X%22%5D%2C%5B1%2C1%2C1%2C1%2C1%2C%22%E2%80%A2%22%2C%22X%22%5D%2C%5B%22X%22%2C1%2C1%2C1%2C1%2C1%2C1%2C%22%E2%80%A2%22%5D%2C%5B%22Measure%22%2C1%2C%22Measure%22%2C1%2C%22Measure%22%2C1%2C%22Measure%22%5D%5D%7D%7D%2C%7B%22id%22%3A%22~95p7%22%2C%22name%22%3A%22Bell%22%2C%22circuit%22%3A%7B%22cols%22%3A%5B%5B%22H%22%5D%2C%5B%22%E2%80%A2%22%2C%22X%22%5D%5D%7D%7D%5D%7D}{Quirk
circuit}. Because the
number of $+1$ outcomes is always even, any single $X$ error is guaranteed to be
detected. Figure~\ref{fig:ghz_correction_and_detection} shows this in the language of ZX
calculus: a single bit-flip error reduces the diagram to zero, so the error is
detected, whereas a pair of bit-flip errors reduces to a single-qubit Pauli
corrections on the GHZ state.

The required corrections can be summarized as follows, and are readily derived
from the ZX diagram. The four Bell pairs define a cycle linking the four sites,
which is manifest in the ZX representation. To recover a proper GHZ state, the
$(-1)$ measurement outcomes are paired up and removed by applying $X$ to the
qubits connecting each pair. The cyclic structure means that the two possible
pairings are equivalent. An odd
number of $(-1)$ outcomes cannot be fully paired, which signals that an error
has occurred.

\begin{figure}[t]
    \centering
    \includegraphics[width=\linewidth]{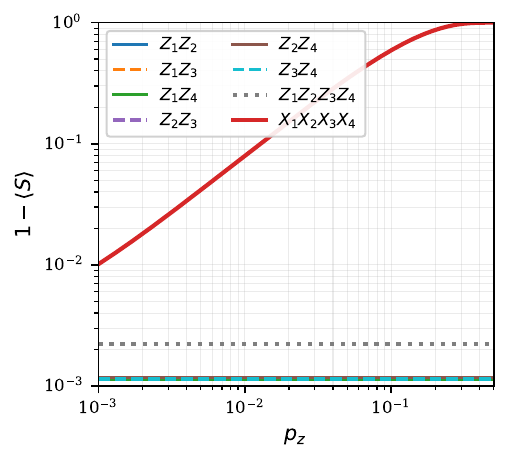}
    \caption{\textbf{Bias preservation in the four-body GHZ state.} Expectation value error of the eight non-trivial stabilizers of the GHZ state $\ket{\Psi}$ as a function of the input Bell pair phase-error rate $p_z$. A GHZ state is being composed from Bell pairs in the process described in fig. \ref{fig:ghz_generation_protocol}(a). The bit-flip rate on the original Bell pairs and residual operation error are fixed at $p_x = \pres=10^{-3}$. All $Z$-type stabilizers ($Z_i Z_j$ and $Z_1 Z_2 Z_3 Z_4$) fidelity is limited only by $ \pres $, since they are only second order sensitive to $p_x$ error due to error detection, and are insensitive to $p_z$. Only the $X_1 X_2 X_3 X_4$ stabilizer degrades with increasing $p_z$, resulting in bulk-measurement-errors. }    \label{fig:ghz4_z_stabilizers}
\end{figure}

Bit-flip ($X$) errors on the Bell pairs propagate as measurement-flip errors,
together with residual $X$ errors on the GHZ state, and are therefore detectable
through the pairing procedure above. Phase ($Z$) errors, by contrast, are never
registered by the measurements, so the protocol increases the bias between $X$
and $Z$ errors. This can be seen in the simulation presented in Fig.~\ref{fig:ghz4_z_stabilizers}, where the $X$ noise on the bell pairs is weak with probability $\pres=10^{-3}$, and the phase noise $p_z$ is scanned but never leaks to the $Z$ stabilizers. 

Even if specifically the $X/Y$ error rate of the Bell pair was higher, we can identify such bit-flip errors and are only second-order sensitive to it. Therefore, this protocol can accommodate much higher noise in this specific direction. $X$ errors on the control qubit of the $CX$ gates is not detectable, as it doesn't propogate to the measured qubit. We threshold include in our noise model a first-order $\pres$ $X$ errors on each qubit in the GHZ state.

\section{Stabilizer code noise model}
\label{app:QEC_noise_model}

In this appendix we describe the noise model used in our surface-code
simulations. We first state the underlying physical assumptions and then
introduce the simplified circuit-level model used in practice.

\begin{figure*}
    \centering
    \includegraphics[width=\textwidth]{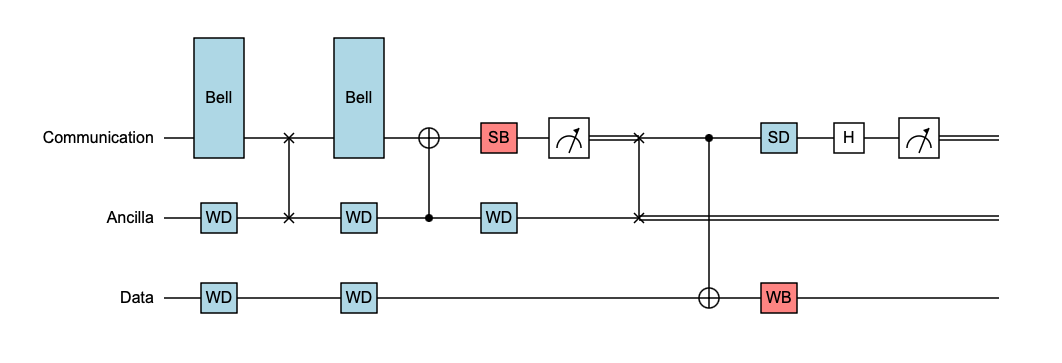}
    \caption{\textbf{Physical noise model for GHZ-based syndrome extraction, from
    the perspective of a single node.} Only the communication qubit (electron
    spin) is optically active, so Bell pairs are generated on it and moved onto
    the ancilla nucleus with a {SWAP}, allowing two Bell pairs to be
    produced in sequence. Box colors indicate the dominant Pauli error of each
    noisy operation: light blue for dephasing ($Z$) and light red for bit flip ($X$). SD/SB
    (strong dephasing/bitflip) and WD/WB (weak dephasing/bitflip) denote the strong electron and
    weak nuclear dephasing during the $ZZ$-based two-qubit gates; red boxes mark
    the cases in which the single-qubit rotations converting $ZZ$ into {CX}
    rotate this dephasing into an $X$ error. The Bell boxes carry a strong $Z$
    and a weak $X/Y$ component. The strong $X$ error preceding the mid-circuit
    measurement is detected and post-selected away. Noise on the {SWAP} is
    neglected here.}
    \label{fig:single_node_ghz_generation_error_model}
\end{figure*}

Figure~\ref{fig:single_node_ghz_generation_error_model} shows GHZ-based syndrome
extraction from the perspective of a single node. We assume that only the
communication qubit (the electron spin) is directly accessible: it can emit
photons to generate Bell pairs, be measured, and undergo two-qubit gates with
the nuclear spins. Consequently, feeding a node into the shared GHZ state
requires generating two Bell pairs in sequence, interleaved with a {SWAP}
that moves the first half-pair onto the ancilla nucleus so that the communication
qubit is free to generate the second. For now we neglect the noise of the
{SWAP}. 

Colored boxes mark noisy operations, with the color indicating the dominant
Pauli term they contribute: light blue for $Z$ (dephasing) errors and light red
for $X$ errors. The two-qubit gates are realized through a $ZZ$ interaction and
are therefore limited by the dephasing of the participating spins. We model this
as a $Z$ error on both the electron and the nucleus, much stronger on the
electron, and label the two contributions SD (strong dephasing) and WD (weak
dephasing). Because the bare $ZZ$ interaction is converted into a {CX} by
single-qubit rotations, on the rotated qubit this dephasing manifests as an $X$
error rather than a $Z$ error; we mark these cases as red boxes. The idling of
the ancilla nucleus during Bell pair generation is included as a weak-dephasing
($Z$) error.

This leaves three dominant noise sources: (i) the Bell pairs, which carry a
strong $Z$ component together with a weak $X/Y$ component, reflecting the bias
produced by the double-click protocol; (ii) a strong $X$ error immediately
before the mid-circuit measurement, which is detected and post-selected away;
and (iii) a strong $Z$ error following the final {CX} that measures the
syndrome. Since the strong $X$ error of (ii) is detectable, the effective model
retains only a strong $Z$ error and weak $X/Y$ errors. These set the two error
scales used throughout the paper: the phase-error rate $p_z$ and the residual
rate $p_\text{res}$. We always assume $p_\text{res}\ll p_z$.
com
The resulting circuit-level model is shown in
Figure~\ref{fig:ghz_noise_model}. Gray boxes denote $p_\text{res}$-scale
noise---data-qubit idling, single-qubit $X/Y$ errors, and two-qubit depolarizing
($D_2$) channels---while a single orange $Z$ marks the $p_z$ phase error on each
ancilla qubit. This term absorbs both the short coherence time of the electron
and the biased Bell pair, and we take it to represent the entire error budget of
the low-coherence qubit. Note that such a $Z$ error commutes through the
{CX} and only flips the measured syndrome, never propagating to the data;
the dominant $X$ component of the GHZ state is second order and is folded into
$p_\text{res}$ (see Appendix~\ref{SI:GHZ}).

Finally, since the data qubit idles during the two Bell pair generation steps,
we apply an idling error of strength $2p_\text{res}$ between rounds. Physically
this noise is biased to pure $Z$,  which would normally align with one type of
stabilizer and leave the code exposed along that axis. This can be mitigated,
however, by using an XZZX surface code~\cite{Bonilla_Ataides_2021}: under the local Hadamard rotation that
defines the XZZX code on half of the qubits, the uniform $Z$ idling is mapped to
$Z$ on one sublattice and $X$ on the other, so that the biased noise is spread
symmetrically across both stabilizer types and appears isotropic to the decoder.
We therefore model the idling directly in this rotated frame, applying a $Z$ or
$X$ error with probability $2p_\text{res}$ to the two sublattices in a
checkerboard pattern.

\section{Hook error protection for high weight stabilizers}
\label{SI:hook_errors}

Hook errors are ancilla errors that occur during the syndrome extraction
circuit and propagate outward to the data qubits. For a syndrome of weight
$w$, such an error can spread to any number of qubits between $1$ and $w$.
However, because of the structure of the syndrome extraction circuit, only
errors of the same Pauli type as the syndrome can propagate outward from the
ancilla, making them equivalent up to the action of the stabilizer. To make
this concrete, suppose the syndrome extraction acts on four qubits labeled
$1,2,3,4$ and an ancilla measures the stabilizer $S = X_1 X_2 X_3 X_4$ in
that order. A noisy ancilla can then produce any of the errors
$X_1 X_2 X_3 X_4$, $X_2 X_3 X_4$, $X_3 X_4$, or $X_4$. Since $S$ is measured,
all of these are equivalent up to application of $S$, so in practice the four
error types can be represented by $I$, $X_2$, $X_3 X_4$, and $X_4$,
which are at most weight-$2$ errors.

In the more general case of a weight-$w$ stabilizer, the same idea applies,
and we can find a representative for each error of weight $\le w/2$. This is
an important issue in practice, since a code of distance $d$ can be reduced to
an effective distance of $2d/w$ by a sufficiently bad syndrome extraction
circuit.

In our setting, using the $w=4$ surface code with post-selected GHZ states,
no single error affects more than one data qubit, so the probability of a
two-data-qubit error scales as $\pres^2$, exactly as in the hook-free case.
In this sense, our syndrome extraction circuit contains no hook mechanism. At
order $p^2$, any pair of data qubits may be corrupted, but this still yields
the full distance of the code. For the surface code specifically, and for any
hypergraph product code \cite{Manes_2025, Tan_2025}, there exist single-ancilla
circuits for which hook errors do not reduce the effective distance. This is
not generally true for other codes.

Our protocol also provides an advantage for codes with $w > 4$. The same
GHZ-generation protocol extends straightforwardly to any weight $w$, retaining
the ability to detect a single $X$ error, so no order-$p$ error propagates to
more than a single data qubit. For two simultaneous errors the situation is
slightly more involved. A pair of errors may simply flip two data qubits,
which does not change the effective circuit distance. Alternatively, the two
errors may flip two of the GHZ measurements, producing a multi-qubit
correction. Such a correction can involve as many as $w-1$ qubits, which, by
the same argument as above, reduces to at most $\lfloor w/2 \rfloor$ errors.

It follows that the circuit distance $d_{\text{circ}}$ satisfies
\begin{equation}
    d_{\text{circ}} \ge \frac{2d}{\lfloor w/2 \rfloor},
\end{equation}
which for $w=4$ recovers the standard distance. As an example, the bivariate
bicycle code \cite{Bravyi_2024} with $w=6$ gives $d_{\text{circ}} \ge 2d/3$,
twice as large as the naive circuit distance $d_{\text{circ}} = d/3$. The
circuit distance of such codes is known to exceed these bounds substantially,
suggesting that with the GHZ advantage it may be possible to recover the full
circuit distance.

\section{Stability Experiment with Bulk measurement error}
\label{SI:stability}

The behavior of lattice surgery is best understood by contrasting memory and
stability experiments. The stability experiment, introduced in~\cite{Gidney_stability_2022}, is a
dual of the standard memory experiment, in which only time-like error chains can
produce a logical fault. It isolates precisely the mechanism by which time-like
errors corrupt a lattice-surgery measurement, while discarding the additional
space-like errors that afflict the static memory qubits making up the surgery
region. Because only sufficiently long time-like chains can cause a logical
fault, lengthening the experiment improves the logical fidelity without bound:
the effective distance is simply the number of measurement rounds.

\begin{figure*}
    \centering
    \includegraphics[width=\textwidth]{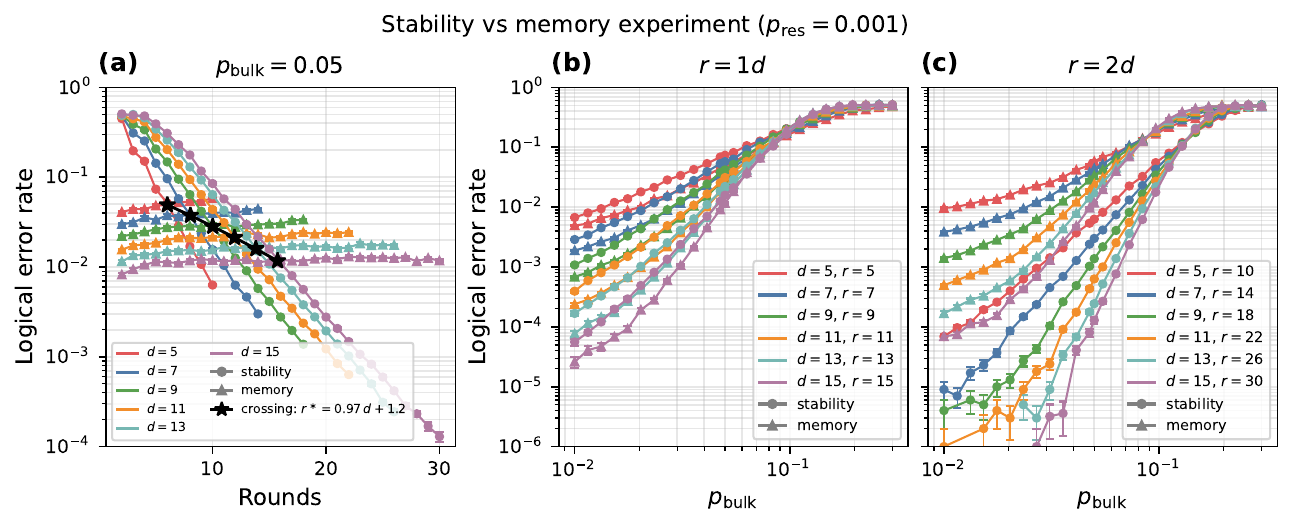}
    \caption{Comparison of the stability and memory experiments under the
    bulk-measurement-error noise model at $p_\text{res}=10^{-3}$, with
    $p_\text{bulk}$ simulated directly (no GHZ ancilla) and syndrome extraction
    implemented with one ancilla per stabilizer using $CZ$ and $H$ gates.
    (a)~Logical error rate versus the number of rounds at fixed
    $p_\text{bulk}=0.05$ for code distances $d=5$--$15$. The stability curves
    (solid) decrease without bound, while the memory curves (dashed) plateau at
    the quantum-memory floor. The two cross near $d$ rounds. (b,c)~Logical error
    rate versus $p_\text{bulk}$ for $r=d$ (b) and $r=2d$ (c). Since stability experiment is more susceptible to bulk errors, if improves faster as $p_\text{bulk}$ is reduced. }
    \label{fig:stability_rounds}
\end{figure*}

Bulk-measurement errors are exactly the time-like errors that drive logical
faults in a stability experiment, but as noted in the main text, their effect
can be suppressed by repeating the measurements. Figure~\ref{fig:stability_rounds}
shows stability and memory experiments in which $p_\text{bulk}$ is simulated
directly, rather than through a GHZ ancilla, with syndrome extraction performed
using a single ancilla per stabilizer and $CZ$ and $H$ gates (with $p_\text{res}=10^{-3}$ fixed). As the number of
rounds is increased, the stability experiment improves indefinitely, whereas the
lattice-surgery simulation is ultimately limited by the quantum-memory
error rate. The two curves cross at approximately $d$ rounds, depending on the ratio of $p_\text{res}$ and $p_\text{GHZ}$.

\section{Effective Error Model for Biased Double-Click Bell pairs}
\label{app:dcp_error_model}

For the sake of completeness we describe the double-click protocol~\cite{barrett2005efficient,Campbell_2008} illustrated in Fig.~\ref{fig:DCP_PROTOCOL}. 
Each node contains two long-lived spin states, \(\ket{\uparrow}\) and \(\ket{\downarrow}\), and an optically excited state \(\ket{e}\). 
The optical field \(\Omega_L(t)\) drives the spin-selective transition
\begin{equation}
\ket{\downarrow}\rightarrow \ket{e},
\qquad
\ket{\uparrow}\rightarrow \ket{\uparrow},
\end{equation}
whereas the microwave field \(\Omega_M(t)\) drives rotations in the ground-state spin manifold,
\begin{equation}
\ket{\uparrow}\leftrightarrow \ket{\downarrow}.
\end{equation}
Spontaneous emission from the excited state returns the system to \(\ket{\downarrow}\) while emitting a photon,
\begin{equation}
\ket{e} \ket 0_{\gamma}\rightarrow \ket{\downarrow}\ket{1}_\gamma .
\end{equation}
The emitted photons leak out of the two cavities, interfere on a balanced beam splitter, and are detected by photo-detectors \(D_+\) or \(D_-\).

The protocol proceeds as follows.

\begin{enumerate}
    \item \textbf{Initialization.}
    A microwave \(\pi/2\)-pulse prepares each spin in
    \begin{equation}
    \ket{p_1}
    =
    \sqrt{1-p_1}\ket{\uparrow}+\sqrt{p_1}\ket{\downarrow} .
    \end{equation}

    \item \textbf{First optical excitation and heralding.}
    An optical \(\pi\)-pulse with \(\Omega_L(t)\) excites the optically active \(\ket{\downarrow}\) state. 
    The beam-splitter output modes are
    \begin{equation}
    c_\pm
    =
    \frac{c_A\pm c_B}{\sqrt{2}},
    \end{equation}
    where \(c_A\) and \(c_B\) denote photon emission from nodes \(A\) and \(B\). 
    A click in detector \(D_\pm\) applies the jump operator \(c_\pm\). 
    Acting on the one-photon sector,
    {\footnotesize $
    c_\pm
    \left(
    \ket{\downarrow\uparrow}\ket{1_A0_B}
    +
    \ket{\uparrow\downarrow}\ket{0_A1_B}
    \right)
    \propto
    \ket{\downarrow\uparrow}
    \pm
    \ket{\uparrow\downarrow}.
    $}
    Thus a single detector click projects the single-emission component onto
    \begin{equation}
    \ket{\Psi_\pm}
    =
    \frac{
    \ket{\downarrow\uparrow}
    \pm
    \ket{\uparrow\downarrow}
    }{\sqrt{2}},
    \end{equation}
    with the sign fixed by the detector outcome.

    \item \textbf{Relaxation after the first click.}
    The first heralding event does not by itself prepare a pure Bell state, because the initial \(\ket{\downarrow\downarrow}\) component contains two optically active spins which will emit two photons. Even with high efficiency photon-number-resolving detection, the overall
    photon collection and detection efficiency is finite. Therefore, a two-photon emission event from the
    \(\ket{\downarrow\downarrow}\) branch can still be registered as a single
click if one photon is lost or missed. the conditional spin state will be
    $
    \rho^{(1)}
    =
    (1-p_{\downarrow\downarrow})\ket{\Psi_{s_1}}\bra{\Psi_{s_1}}
    +
p_{\downarrow\downarrow}\ket{\downarrow\downarrow}\bra{\downarrow\downarrow},
    $
where \(s_1=\pm\) is determined by the first detector click and \(p_{\downarrow\downarrow}\) is determined by the efficiency of resolving one and two photons as well as the loss rate as loss of a single photon on the way will result in a single click. If two photons are emitted and each is detected with probability \(\eta_{\rm ph}\), the probability that this two-emission event is recorded as exactly one detected photon is \(2\eta_{\rm ph}(1-\eta_{\rm ph})\). Therefore, in a detection-only estimate with similar click-time densities for the one- and two-emission branches,
{\footnotesize \begin{align}
p_{\downarrow\downarrow}
&=\frac{p^{2}_{1}\eta_{{\rm ph}}(1-\eta_{{\rm ph}})}{p_{1}\left(1-p_{1}\right)\,\eta_{{\rm ph}}+p^{2}_{1}\eta_{{\rm ph}}(1-\eta_{{\rm ph}})}\\
&=\frac{1-\eta_{{\rm ph}}}{1/p_{1}-\eta_{{\rm ph}}} \nonumber
\end{align}}

In the full trajectory description \cite{barrett2005efficient}, this estimate also depends on the click time through the cavity emission amplitudes.
    \item \textbf{Microwave inversion.}
    A global microwave \(\pi\)-pulse with \(\Omega_M(t)\) flips both spins,
    it preserves the Bell sector up to a phase,
    \begin{align}
    X_A X_B\ket{\Psi_+}
    &=
    \ket{\Psi_+},
    \\
    X_A X_B\ket{\Psi_-}
    &=
    -
    \ket{\Psi_-},
    \end{align}
    while mapping the unwanted same-spin component to
    \begin{equation}
    X_A X_B\ket{\downarrow\downarrow}
    =
    \ket{\uparrow\uparrow}.
    \end{equation}

    \item \textbf{Second optical excitation and heralding.}
    A second optical \(\pi\)-pulse is applied. 
    The mapped same-spin branch \(\ket{\uparrow\uparrow}\) is dark and therefore cannot generate a second heralding click. 
    In contrast, the Bell component remains in the single-excitation sector and can emit one photon. 
    Conditioning on exactly one detector click in the same detector in the second round therefore removes the natural \(\ket{\downarrow\downarrow}\) branch and accepts only the Bell component.

\end{enumerate}

\begin{figure*}[ht!]
   \centering
   \includegraphics[width=1\linewidth]{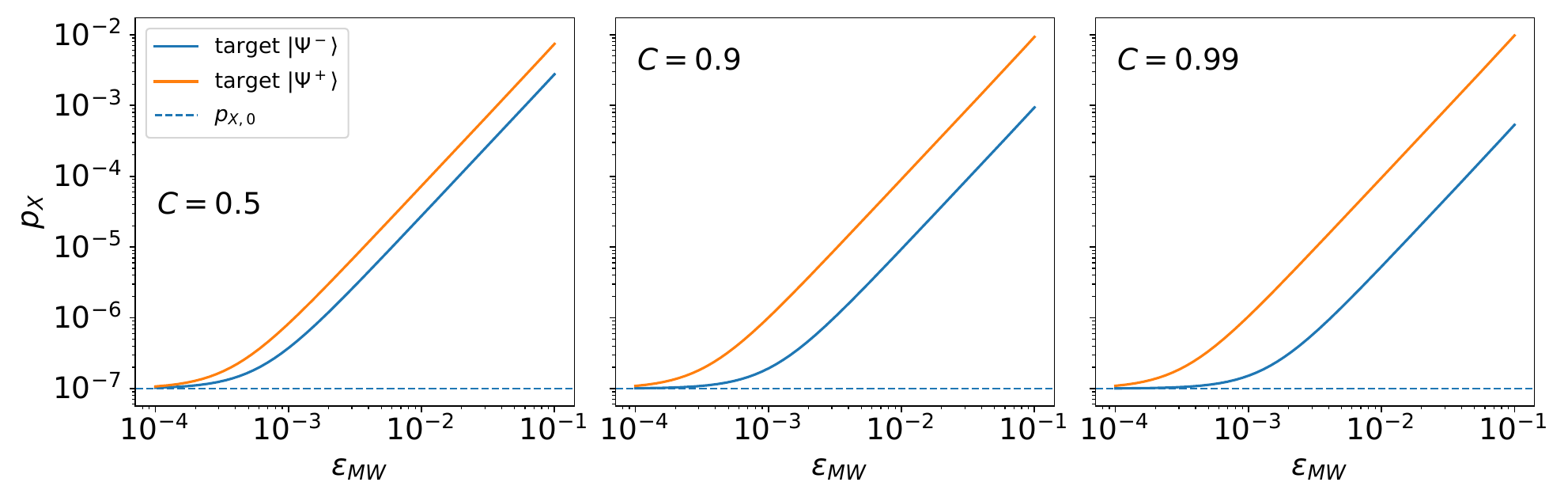}
    \caption{
Comparison of microwave-induced \(X\)-type errors for singlet and triplet Bell pair frames in the DCP,
given correlation \(C=0.50,0.90,0.99\) between the the two MW pulse errors to each qubit .
The singlet target \(|\Psi^-\rangle\) is less sensitive to the common part of the pulse error, while the triplet target \(|\Psi^+\rangle\) is sensitive to it.
As the noise becomes more correlated, the singlet error approaches the accepted floor \(p_{X,0}\), faster. 
}
   \label{fig:dcp_mw_singlet_triplet}
\end{figure*}

An attempt is accepted iff exactly one detector click is recorded in each optical round.

We describe the accepted Bell pair using the phase-error model described in \cite{debone2024thresholds}. In this model the accepted state is Bell diagonal in the \(\{|\Psi^+\rangle,|\Psi^-\rangle\}\) subspace,
\[ 
\rho_{\rm link} = (1-p_Z)|\Psi^+\rangle\langle\Psi^+| + p_Z|\Psi^-\rangle\langle\Psi^-| , 
\] 
so that the link fidelity is identified with a phase-type Bell error, \(p_Z=1-F_{\rm link}\). We model the leading accepted \(X\)-type contribution as coming from the microwave inversion pulse between the two optical rounds. This error depends strongly on the Bell frame. A common microwave pulse-area error is suppressed for the singlet \(|\Psi^-\rangle\), while the triplet \(|\Psi^+\rangle\) leaks into the $\left|\Phi_{\pm}\right>=\left|00\right>\pm\left|11\right>$ sector. Therefore, post selecting on $\ket{\Psi^-}$ introduces an additional path to increase the X/Z bias. 

We parametrize the amplitude of the microwave pulse-area error by \(\epsilon_{\rm MW}\), and the degree to which the two spins experience the same pulse error by a correlation parameter \(C\), with \(C=1\) corresponding to a perfectly common error.

For a triplet-frame target, the leading microwave contribution is 
\[ 
p_{X,{\rm MW}}^{\rm triplet} = (1-p_Z) \frac{\epsilon_{\rm MW}^2}{2}(1+C) + p_Z \frac{\epsilon_{\rm MW}^2}{2}(1-C). 
\] 
This expression makes explicit why the triplet frame remains sensitive to common-mode microwave noise (the \(1+C\) term), unlike the singlet frame which suppresses it. We therefore focus on the triplet frame to provide a conservative estimate of the accepted \(X\)-type error. We show numerical simulation of this relation in Fig.~\ref{fig:dcp_mw_singlet_triplet}.

In addition to the microwave contribution, we include a small false-heralded \(|\uparrow\uparrow\rangle\) branch, which we label by the probability $p_{\uparrow\uparrow}$. This branch is important because the microwave inversion maps it to \(|\downarrow\downarrow\rangle\), where both spins can be optically active in the second optical round. 
The natural \(|\downarrow\downarrow\rangle\) branch is treated separately, since it is removed by the second round and contributes only at higher order in the microwave pulse-area error. 
The total accepted \(X\)-type error in the triplet frame is thus modeled as 
\[ 
p_X^{\rm triplet} = p_{X,{\rm MW}}^{\rm triplet} + p_{\downarrow\downarrow} \frac{\epsilon_{\rm MW}^4}{16} + p_{\uparrow\uparrow}.
\]

\bibliographystyle{apsrev4-1}
\bibliography{bib}
\end{document}